\documentclass{article}
\usepackage{PRIMEarxiv}

\usepackage[utf8]{inputenc} 
\usepackage[T1]{fontenc}    
\usepackage{hyperref}       
\usepackage{url}            
\usepackage{booktabs}       
\usepackage{amsfonts}       
\usepackage{nicefrac}       
\usepackage{microtype}      
\usepackage{lipsum}
\usepackage{fancyhdr}       
\usepackage{graphicx}       
\graphicspath{{media/}}     
\usepackage{tabularx}
\usepackage{dcolumn}
\usepackage{multirow}
\newcolumntype{d}[1]{D{.}{.}{#1}}

\title{What Makes an Apology More Effective? Exploring Anthropomorphism, Individual Differences, and Emotion in Human-Automation Trust Repair
\thanks{\textit{\underline{Citation}}: 
\textbf{Authors. Title. Pages.... DOI:}} 
}

\author{
  Peggy Pei-Ying Lu \\
  Institute for Creative Integration \\
  Oakland, CA, U.S.A.\\
  \texttt{peggy.lu@creative-integration.com} \\
  \And
  Makoto Konishi\\
  Vision Design Div.  \\
  TOYOTA MOTOR CORPORATION \\
  Toyota, Aichi, JAPAN\\
  \texttt{makoto\_konishi@mail.toyota.co.jp} \\
  \AND
  Shin Sano \\
  Institute for Creative Integration \\
  Oakland, CA, U.S.A. \\
  \texttt{ssano@creative-integration.com} \\
  \And
  Sho Hiruta \\
  Institute for Creative Integration \\
  Oakland, CA, U.S.A. \\
  \texttt{shiruta@creative-integration.com} \\
  \And
  Francis Ken Nakagawa \\
  Institute for Creative Integration \\
  Oakland, CA, U.S.A. \\
  \texttt{fnakagawa@creative-integration.com} \\
}

\begin{document}
\maketitle

\begin{abstract}
Recent advances in technology have allowed an automation system to recognize its errors and repair trust more actively than ever. While previous research has called for further studies of different human factors and design features, their effect on human-automation trust repair scenarios remains unknown, especially concerning emotions. This paper seeks to fill such gaps by investigating the impact of anthropomorphism, users’ individual differences, and emotional responses on human-automation trust repair. Our experiment manipulated various types of trust violations and apology messages with different emotionally expressive anthropomorphic cues. While no significant effect from the different apology representations was found, our participants displayed polarizing attitudes toward the anthropomorphic cues. We also found that (1) some personality traits, such as openness and conscientiousness, negatively correlate with the effectiveness of the apology messages, and (2) a person's emotional response toward a trust violation positively correlates with the effectiveness of the apology messages.
\end{abstract}


\section{Introduction}
Trust has an essential role in both human-human interaction and human-automation interaction, and it can be understood as an “\textit{attitude that an agent will help achieve an individual’s goals in a situation characterized by uncertainty and vulnerability.}”\cite{Lee_See_2004}. Early human-automation research has confirmed that automation errors can negatively affect trust and that it is crucial to repair trust when it is damaged to foster a long-term relationship \cite{Merritt_2008} . Moreover, recent developments in artificial intelligence (AI) has allowed an automation system to actively repair trust more than ever.

Considering the development of self-driving vehicles and the increasing adoption of in-car voice agents, trust is especially relevant to the automotive industry. Among the various strategies, apologies have become the most common strategy for an automation system to repair trust today \cite{deVisser_2018}. However, what makes an apology message from an automation system more effective than other strategies? Or maybe more importantly, in the age of personalization, what makes an apology message from an automation system more effective for one individual than another individual? 

Prior studies have ascertained the individuality in trust and acknowledged that a more thorough understanding of the impact of various factors contributing to such individuality is needed \cite{Hoff_2015} \cite{Schaefer_2016} \cite{Kraus_2020_a}. In addition to personality, a person’s emotion has also been identified as a potential contributor to individual differences \cite{Stokes_2010} \cite{Hoff_2015}. Coupling these factors with the growing incorporation of anthropomorphic AI, which has recently demonstrated its ability to express emotions \cite{Goya-Martinez_2016}, we believe further studies focusing on emotion are critical in understanding the complex individuality of human-automation trust.

Therefore, this paper starts by reviewing relevant work in the domain of trust repair in human-automation interaction, specifically regarding individual differences, the role of emotion in the trust repair process, and how anthropomorphic AI may affect the effectiveness of trust repair strategies. We then introduce our 3 (criticality of trust violations) × 3 (apology representations with emotionally expressive anthropomorphic cues) experiment situated within common driving scenarios. Our findings and recommendations for future studies regarding personality and emotion in the context of human-automation trust repair are discussed at the end of the paper.

\section{Related Works}
\subsection{Trust Repair in Human-Automation Interaction}
While there has been extensive research on trust between human and automation, trust repair has only recently gained scholarly attention in human-automation research. De Visser et al. \cite{deVisser_2018} called for opportunities for an automation system to actively build and repair trust with a human-centered approach. Baker et al. \cite{Baker_2018} echoed these views and emphasized the increasing need for a better understanding of human-robot trust repair.

A few empirical studies in recent years have examined various factors relevant for human-automation trust repair. Some studies have replicated previous findings in human-human research regarding the effectiveness of various trust repair strategies addressing different trust violations. \cite{Kim_2004}. For example, Kohn et al.\cite{Kohn_2018} explored a wide range of trust repair strategies with self-driving vehicles and found that some apologies are more effective than other strategies. Quinn \cite{Quinn_2018} examined whether the effectiveness of a trust repair strategy can vary across different violation types and learned that apologies could repair trust harmed by competence-based failures, but in turn, would damage trust harmed by integrity-based failures. 

Similar to human-human trust repair research \cite{Kim_2006}, causal attribution has also been one of the common themes in human-automation trust repair studies. Buchholz et al. \cite{Buchholz_2017} replicated prior human-robot trust repair research and found that a self-blame agent was more trustworthy among cooperative human-agent interactions. Jensen et al. \cite{Jensen_2019} produced a similar result and found that attributing the blame to a system’s developers resulted in a negative perception of the system’s trustworthiness.

An automation system’s human-likeness has also received some attention in recent empirical studies of trust repair. The research we explored in this regard will be discussed in a later section focusing on anthropomorphism.

\subsection{Individual Differences in Human-Automation Trust}
The abovementioned studies and prior human-human trust repair studies have made it evident that no single trust repair strategy should be generically applied, especially considering interpersonal differences \cite{deVisser_2018}. Previous reviews have identified several human traits that can affect how a person perceives and interacts with automation systems in addition to demographic characteristics such as age, gender, and culture. Personality and a person’s propensity to trust also have a substantial role in human-automation trust calibration \cite{Hoff_2015} \cite{Schaefer_2016}. 

In one of the earliest studies investigating personality’s impact on human-automation trust, Merritt and Ilgen \cite{Merritt_2008} found that extraversion positively correlated to a person’s propensity to trust machines. However, subsequent empirical studies have uncovered inconsistent results. For instance, Miramontes et al. \cite{Miramontes_2015} found that air traffic controllers in training with high emotional stability are more likely to trust a new automation system. In terms of general trust in automation, Chien et al. \cite{Chien_2016} reported that only agreeableness and conscientiousness displayed a positive correlation with trust. Rossi et al. \cite{Rossi_2018} found a similar result that higher agreeableness and conscientiousness value can increase a person’s tendency to trust a robot in an emergency scenario. 

It is worth noting that these studies were conducted in different contexts, investigated different types of automation, and even measured different types of trust. Moreover, none of them addressed scenarios in which a trust violation occurs, and thus, did not measure the impact in the trust repair context.

\subsection{Role of Emotion in Trust Repair}
A person’s emotional state is another prominent factor contributing to the individual differences in trust in both humans and automation on a situational level. Early trust researchers have acknowledged the affective nature of trust and argued that emotional reactions are critical to trust \cite{Lee_See_2004}. However, previous reviews of human-human trust repair literature have found that while there had been callings for acknowledging emotion’s effect in repairing damaged trust, few researchers have considered its role empirically \cite{Lockey_2017} \cite{Bagdasarov_2019}. This finding is consistent with our reviews of human-automation trust researches that shows a similar lack of investigation into emotion’s role in the trust repair process. 

As one of the earliest trust models that included emotion as a component, Tomlinson and Mayer \cite{Tomlinson_2009} proposed that if negative emotions can be mitigated by the trust repair strategy, trust is more likely to be repaired. A few experiments in the interpersonal or organizational trust domain have found evidence in a similar vein. Specifically, Lockey \cite{Lockey_2017} found that negative emotion may be influential throughout the whole trust process in certain contexts and that anger and contempt were more influential than fear.

While some human-automation trust studies included emotion as a variable, few have employed a \textit{repair} context. For instance, prior studies have suggested that a person’s initial emotional state impacts their initial trust toward automation; in particular, a positive mood may lead to a higher trust level \cite{Stokes_2010} \cite{Merritt_2011}. On the other hand, Kraus et al. \cite{Kraus_2020_b} found that anxiety predicts trust better than other positive and negative affects when introduced to a new automated driving system. However, it remains unclear whether these effects would also be present in a trust repair scenario.

\subsection{Anthropomorphic AI, Trust, and Emotion}
Anthropomorphism is the act of attributing human-like capacities, such as intention, emotion, and cognition, to non-human agents \cite{Waytz_2010}. Consistent with the computers-are-social-actors framework (CASA), anthropomorphism is considered “\textit{a key determinant of how media agents are evaluated}” \cite{Gambino_2020}. As previously mentioned, anthropomorphism has been a focus of recent human-automation trust literature, and most researchers have found that it has a positive impact on trust \cite{Waytz_2014} \cite{Kulms_2019} \cite{Song_2020}. In the trust repair context, de Visser et al. \cite{deVisser_2016} found that anthropomorphism can dampen the impact of trust violations and improve trust repair in an automation-teaming context.

A recent review of cross-disciplinary studies in trust repair literature has revealed that anthropomorphism is a critical factor affecting emotional trust in virtual AI \cite{Glikson_2020}. Moreover, certain manipulation in attractiveness and anthropomorphic cues have been found to induce positive emotion and increase perceived trustworthiness. For instance, Verbene et al. \cite{Verberne_2015} discovered that the similarity in facial features between a virtual driver and a human driver positively correlates with trust and likeness. Matsui and Yamada \cite{Matsui_2019} found that a virtual agent with expressive facial movements has higher perceived trustworthiness.

\section{Current Study}
This study speculates that the effectiveness of a trust repair attempt can vary depending on whether the trust repair representation features any anthropomorphic cues that convey emotion. We also examine whether a participant’s individual differences and emotional responses to a trust violation may impact the effectiveness of a trust repair strategy. Therefore, in creating this experiment, we formulated the following hypotheses: 

\begin{itemize}
\item H1: An apology message featuring anthropomorphic cues that are emotionally expressive will impact how effective a trust repair strategy is.
\item H2: A participant’s personality impacts the effectiveness of a trust repair strategy.
\item H3: The participant’s post-trust-violation emotional response will impact the effectiveness of a trust repair strategy.
\end{itemize}

The goal of this study is twofold. First, we would like to fill in the gap of previous research by exploring factors that have been identified as relevant in influencing trust in a more specific context, i.e., trust repair. To the best of our knowledge, this study is the first human-automation trust repair study that attempted to explore the role of emotion in the process of trust repair. Second, we hope that the result will not only provide us with a more focused next step but also inform and inspire future researchers and practitioners working in the same area.

\section{Methods}
\subsection{Design}
We employed a 3 × 3 repeated measures design, with criticalities of trust violations (\textit{low}, \textit{mid}, and \textit{high}) being the within-subject factor and different system apology representations (\textit{text}, \textit{emoji}, and \textit{cartoon car}) the between-subject factor. All participants received three trust violation scenarios and one apology representation at the end of each scenario. The order of the trust violation scenarios was fully counterbalanced, while the apology representation was randomly assigned.

A few conditions were controlled to avoid known impacts from other factors that were beyond the scope of this study. Specifically, all trust violation scenarios presented in this study are competency-based errors. Moreover, all trust repair strategies are apologies with internal blame attribution. Prior research has revealed that such a pair between a trust violation and a trust repair strategy are the most common and effective in human-automation interaction \cite{Quinn_2018} \cite{Jensen_2019}. Furthermore, we acknowledged that trust might be affected by error types, as indicated by prior research \cite{Kohn_2018} \cite{Mishler_2019}. Thus, a few trust violation scenarios featuring a different criticality of errors were included in the experiment.

Additionally, in this study, we are interested in understanding trust repair in the context that there was a pre-existing relationship between the user and the system. Hence, we chose to situate the trust violation scenarios in today’s common driving scenarios with which most participants would be familiar and likely have some personal experiences instead of introducing a new system to them. This setup is also devised in response to critics of Hoff and Bashir \cite{Hoff_2015}: that among the studies they reviewed that manipulated design features, most utilized very specialized and goal-oriented tasks that were far removed from most people’s day-to-day experiences.

\subsection{Participants}
A total of 1269 participants were recruited through SurveyMonkey, an online survey platform. The study was restricted to respondents older than 18 who originated in the United States of America. Participants who did not pass the screening question or did not complete the whole survey were eliminated from the results. 

Among the remaining 1074 participants, 52\% of the participants were female. Regarding age, 24\% of the participants were aged 18 to 29, 26\% were aged 30 to 44, 29\% were aged 45 to 60, and 21\% were older than 60. The distribution of the age group should be proportional to the population distribution according to the United States of America census through SurveyMonkey’s recruitment mechanism.  

Table~\ref{tab:participants} shows the number of participants who received each apology representation in each trust violation scenario. Chi-squared analyses revealed no significant $(p > .05)$ differences among groups (criticalities of trust violations × apology representations) with respect to gender and age.

\begin{table}[h]
    \caption{Number of participants by criticalities of trust violations × apology representations}
    \label{tab:participants}
    \begin{tabular}{lcccl}
        \toprule
         & Low-criticality & Mid-criticality & High-criticality & Total\\
        \midrule
        \textit{Text} & 344 & 339 & 347 & 1030\\
        \textit{Emoji} & 378 & 376 & 353 & 1107\\
        \textit{Cartoon car} & 352 & 359 & 374 & 1085\\
        \bottomrule
    \end{tabular}
\end{table}

\subsection{Materials}
The different criticality of trust violation scenarios (\textit{low}, \textit{mid}, and \textit{high}) was manipulated by vignettes describing common driving scenarios, centering on different criticality of errors committed by an in-car Global Positioning System (GPS) navigation system. A brief description of the scenario utilized at each criticality level is shown in Table~\ref{tab:violations}. Each scenario was approximately 100 words in length and featured 3 to 4 images between paragraphs as stimuli to elicit participants’ cognitive and affective responses to the scenarios. 

\begin{table}[h]
    \caption{Types of trust violations used in this study}
    \label{tab:violations}
    \begin{tabularx}{\linewidth}{lX}
        \hline
        \toprule
        Level of Criticality & Scenario Descriptions \\
        \midrule
        Low-criticality & {The navigation system provided a detour with no apparent reason resulting in a 5–minute delay.} \\
        Mid-criticality & {The navigation system failed to provide accurate traffic information resulting in the person being late for work.}\\
        High-criticality & {The navigation system arrived at the wrong destination point resulting in the person being stranded in the middle of a mountain area.}\\
        \bottomrule
    \end{tabularx}
\end{table}

\begin{figure}[h]
    \includegraphics[width=\columnwidth]{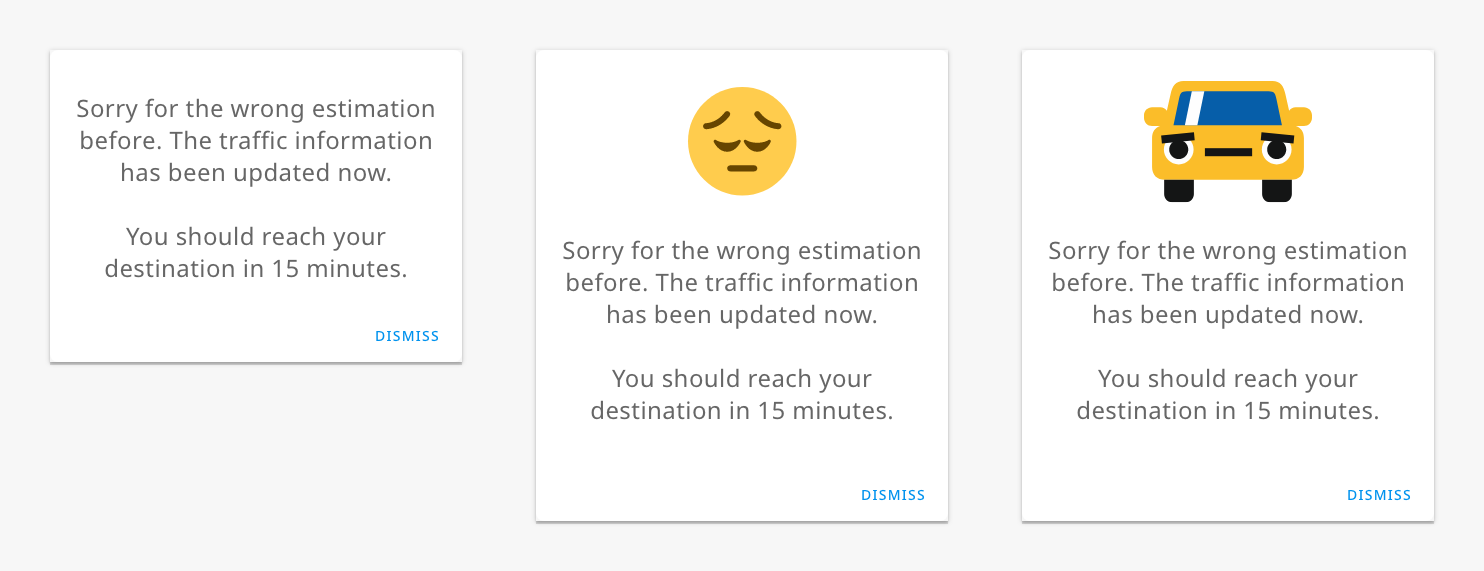}
    \caption{The apology representations (\textit{text}, \textit{emoji}, and \textit{cartoon car}) used in this study addressing the mid-criticality trust violation scenario.}
    \label{fig:apologyRepsMid}
\end{figure}

The different apology representations (\textit{text}, \textit{emoji}, and \textit{cartoon car}) were manipulated by static mock-ups of the user interface. They differed in whether the apology message contained any anthropomorphic cues, specifically, some elements that were human-like features conveying emotions. Each trust violation scenario was paired with its set of apology representations that were appropriate for addressing the specific system error presented in the scenario.\textit{Text} displayed the apology message in plain text, whereas \textit{emoji} and \textit{cartoon car} showed respective visuals expressing an apologetic emotion in addition to the same apology message in \textit{text}. The set of apology representations applied for the mid-criticality trust violation scenario is provided in Figure~\ref{fig:apologyRepsMid} as an example. 

\subsection{Measures}
\subsubsection{Trust Repair}
In this study, we define the effectiveness of the trust repair strategy as the change in the system’s perceived trustworthiness from (a) before a trust violation occurs and (b) after a trust repair strategy is presented. Both (a) and (b) were measured using one subjective question adapted from prior studies \cite{Lee_Moray_2004} \cite{Quinn_2018}:  

\begin{samepage}
    \renewcommand{\labelenumi}{(\alph{enumi})}
    \begin{enumerate}
    \item Reflecting on your experiences using the navigation system that you use most frequently, to what extent do you trust it in general?
    \item If this happens to the navigation system you used most frequently, to what extent will you trust it next time?
    \end{enumerate}
\end{samepage}

Participants answered both questions using a 0–100 visual analog scale, where higher points indicated higher perceived trustworthiness. We then subtracted (a) from (b) to calculate the effectiveness of the presented trust repair strategy. A higher value indicated that the trust repair strategy is more effective. 

\begin{figure}[h]
    \includegraphics[width=245pt]{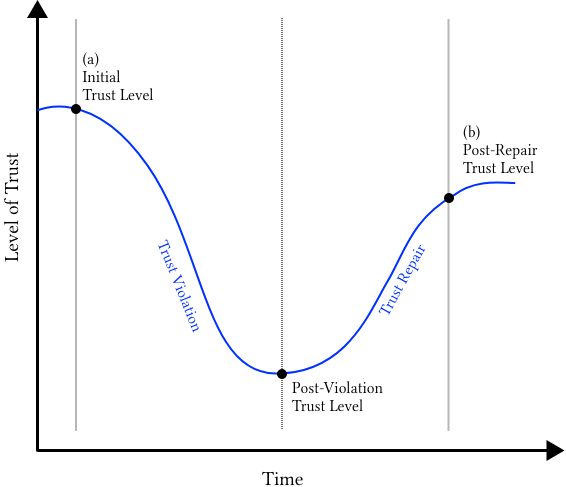}
    \caption{Illustration of trust repair cycle adapted from Lewicki and Brinsfield \cite{Lewicki_2017}}
    \label{fig:trustCycle}
\end{figure}

This relative calculation of change in a person’s trust level was in response to Lewicki and Brinsfield's \cite{Lewicki_2017} call for considering the pre-existing level of trust in their proposed trust violation and repair cycle, as shown in Figure~\ref{fig:trustCycle}. 

\subsubsection{Individual Differences}
In addition to age group and gender demographics, we also measured participants’ personality and trust stance. Following prior studies \cite{Rossi_2018}, the Ten Item Personality Measure (TIPI) \cite{Gosling_2003} was employed to examine differences in personality traits. Regarding the trust stance, we applied a 3-item scale developed by McKnight et al. \cite{McKnight_2002}, as shown in Table~\ref{tab:trustStanceQ}. Participants answered each question under this section using a 7-point Likert scale, ranging from \textit{disagree strongly} to \textit{agree strongly}.   

\begin{table}[h]
    \caption{Questionnaire used to measure trust stance}
    \label{tab:trustStanceQ}
    \begin{tabularx}{\linewidth}{lX}
        \toprule
        & Trust Stance Scale\\
        \midrule
        Q1 & I usually trust people until they give me a reason not to trust them.\\
        Q2 & I usually give people the benefit of the doubt when I first meet them. \\
        Q3 & My typical approach is to trust new acquaintances until they prove I should not trust them.\\
        \bottomrule
    \end{tabularx}
\end{table}

\subsubsection{Emotional Response}
We measured participants’ emotional responses using a two-item questionnaire adapted from Positive and Negative Affect Schedule (PANAS) \cite{Watson_1998} and Emotrak \cite{Garcia_2016}. At the end of each trust violation scenario, participants completed the subjective evaluation of their emotional response both before and after they were presented with the apology representation. Each time, participants chose one emotion from a range of five emotions (\textit{anxious}, \textit{frustrated}, \textit{indifferent}, \textit{satisfied}, and \textit{relieved}) and then rated the intensity of the selected emotion using a visual analog scale of 1 to 5. 

Each response was assigned a score from -5 to 5 based on the chosen emotion and rated intensity for analysis. The score would be equal to the rated intensity if the participant had chosen a positive emotion (\textit{satisfied} or \textit{relieved}). On the other hand, if the participant had chosen a negative emotion (\textit{anxious} or \textit{frustrated}), the score would be their rated intensity times -1. For example, if a participant indicated that they felt \textit{frustrated} and rated the intensity as 2, the assigned score would be -2. Participants chose \textit{indifferent} would be assigned a score of 0 regardless of their rated intensity.

\subsection{Procedures}
The survey started with a brief introduction, and participants first answered a screening question about their commonly used navigation system when they usually drive. They were also asked to rate their trust level toward said navigation system. Participants who had not driven with any navigation system in the last 30 days were disqualified from the study.

Participants then completed a measurement of their individual differences and were introduced to the trust violation scenarios. During the introduction, participants were explicitly prompted to relate the scenarios to the navigation system that they used most often to base their cognitive and affective responses. Each participant completed three trust violation scenarios with the orders fully counterbalanced across participants. They also received instruction about viewing the three scenarios as independent events.

At the end of each scenario, the participants were asked to answer their post-trust-violation emotional response questions. One of the three apology representations (\textit{text}, \textit{emoji}, and \textit{cartoon car}) was randomly assigned afterward. Upon seeing the apology message, the participants were asked about their post-repair emotional response and perceived trustworthiness of the navigation system again. An optional free response then prompted the participants to explain their answers regarding perceived trustworthiness and emotional responses. The whole survey took approximately 5–8 minutes to complete.

\section{Results}
\subsection{Effectiveness of the Apology Representations}
A 3 (violation type: \textit{low}, \textit{mid}, and \textit{high} criticality) × 3 (apology representation: \textit{text}, \textit{emoji}, and \textit{cartoon car}) mixed repeated-measures analysis of variance (ANOVA) revealed a significant main effect of violation type $(F(2, 3219) = 51.15, p < 0.001, \eta^2 = .03)$. The main effect of the apology representations and the interaction between violation type and the apology representations were, however, found to be non-significant for this study. A post hoc Tukey honestly significant difference (HSD) test showed that there is a significant difference among the three trust violation scenarios in the following direction: $\textit{low}  (M = -20.27, SD = 23.58) > \textit{mid}  (M = -24.16, SD = 26.19) > \textit{high}  (M = -31.25, SD = 26.65)$.

\subsection{Individual Differences in Effectiveness of the Apology Representations}
\subsubsection{Personality}
The Pearson correlation analyses showed that only two dimensions—\textit{Openness} and \textit{Conscientiousness}—consistently and significantly correlated to the effectiveness of the apology representations for all three criticalities of the trust violation scenarios. The correlation coefficients and their respective statistical significance are shown in Table~\ref{tab:psnlCor}. The results indicated that the higher is the Openness or Conscientiousness in a person’s personality traits, the less effective is the trust repair strategies that we presented in this study. The result also suggests that the more critical the trust violation is, the stronger the correlation is. 

\subsubsection{Trust Stance}
The Pearson correlation analyses showed that trust stance consistently and significantly correlated to the effectiveness of the apology representations for all three criticalities of the trust violation scenarios. The correlation coefficients and their respective statistical significance are shown in Table~\ref{tab:psnlCor}. The results showed a negative correlation. However, the correlations are relatively weak under all criticalities of the trust violation scenarios.

\begin{table}[h]
    \caption{Correlations between effectiveness of the apology representations and individual differences}
    \label{tab:psnlCor}
    \begin{tabular}{ld{-1}d{-1}d{-1}}
        \toprule
        & \multicolumn{1}{c}{Low-criticality} 
        & \multicolumn{1}{c}{Mid-criticality} 
        & \multicolumn{1}{c}{High-criticality}\\
        \midrule
        Openness            & -0.065^*      &-0.117^{***}  &-0.124^{***} \\
        Conscientiousness   & -0.094^{***}  &-0.153^{***}  &-0.203^{***} \\
        Extraversion        & -0.048        &-0.024        &-0.011 \\
        Agreeableness       & 0.007         &-0.075^*      &-0.045 \\
        Neuroticism         & -0.005        &0.004         &-0.004 \\
        Trust Stance        & -0.059^*      &-0.072^*      &-0.069^*\\
        \bottomrule
        \multicolumn{4}{l}{Significance codes: $ ^* p < 0.05, ^{**} p < 0.01, ^{***} p < 0.001$}
    \end{tabular}
\end{table}

\subsubsection{Age and Gender}
A 4 (age group: \textit{18–29}, \textit{30–44}, \textit{45–60}, and \textit{>60}) × 3 (apology representation: \textit{text}, \textit{emoji}, and \textit{cartoon car}) mixed repeated-measures ANOVA revealed a significant main effect of age group for effectiveness of the apology representations (\textit{low-criticality}: $F(4, 1069) = 2.81, p < 0.05, \eta^2 = .01$; \textit{mid-criticality}: $F(4, 1069) = 3.98, p < 0.01, \eta^2 = .01$; \textit{high-criticality}: $F(4, 1069) = 6.78, p < 0.001, \eta^2 = .02$). The interaction between the age group and the apology representations was, however, found to be non-significant. Descriptive statistics for participant by age group and criticalities of trust violations are shown in Table~\ref{tab:ageDesc}. 

\begin{table}[h]
    \caption{Effectiveness of the apology representations for each participant group by criticalities of trust violations × age group}
    \label{tab:ageDesc}
    \begin{tabular}{lcccc}
        \toprule
         \multirow{2}{*}{Age Group} & \multirow{2}{*}{$N$} 
         & \multicolumn{3}{c}{$M(SD)$} \\
         \cmidrule(r){3-5}
          & & Low-criticality & Mid-criticality & High-criticality\\
        \midrule
        18–29 & 263 & $-18.07 (23.14)$ & $-20.38 (28.13)$ & $-27.67 (26.71)$\\
        30–44 & 278 & $-20.36 (24.33)$ & $-24.29 (24.73)$ & $-30.59 (26.28)$\\
        45–60 & 308 & $-19.13 (23.53)$ & $-23.46 (26.04)$ & $-29.89 (26.77)$\\
        >60   & 225 & $-24.15 (22.69)$ & $-29.22 (25.06)$ & $-37.75 (25.60)$\\
        \bottomrule
    \end{tabular}
\end{table}

Table \ref{tab:ageTukey} shows the post hoc comparisons of the effectiveness of the apology representations using the Tukey HSD test. The results revealed that the effectiveness of the apology representations was significantly higher in the age group \textit{18–29} than \textit{>60} across all criticality levels. In particular, under the high criticality scenario, there are significant differences between the age group \textit{>60} and all other age groups, among which the apology messages were least effective for them, especially between age group \textit{>60} and the age group \textit{18–29}. Overall, the results suggest that there are generational differences in the effectiveness of the apology messages. In particular, similar to the effects of personality, our results suggest that these differences are more substantial when the trust violation is more critical.

\begin{table}[h]
    \caption{Post hoc comparisons among different age groups}
    \label{tab:ageTukey}
    \begin{tabular}{lld{-1}d{-1}d{-1}}
        \toprule
        \multicolumn{2}{c}{\multirow{2}{*}{Age Group}}
        & \multicolumn{3}{c}{Mean Difference}\\
        \cmidrule(r){3-5}
        & & \multicolumn{1}{c}{Low-criticality} 
        & \multicolumn{1}{c}{Mid-criticality} 
        & \multicolumn{1}{c}{High-criticality}\\
        \midrule
        \multirow{3}{*}{>60} & 18–29 & -6.08^* & -8.84^* & -10.07^{***}\\
        & 30–44 & n.s. & n.s. & -7.15^*\\
        & 45–60 & n.s. & n.s. & -7.85^{**}\\
        \bottomrule
        \multicolumn{4}{l}{Significance codes: $ ^* p < 0.05, ^{**} p < 0.01, ^{***} p < 0.001$}
    \end{tabular}
\end{table}

Regarding gender, the T-test analyses did not indicate any statistically significant differences in their trust change for the three criticalities of scenarios.

\subsection{Emotional Response and Effectiveness of the Apology Representations}
The Pearson correlation analyses between participants’ post-trust-violation emotional response and the effectiveness of the apology representations revealed a positive correlation that is statistically significant for all three levels of criticality. Refer to Table~\ref{tab:emoCorr} for their respective correlation coefficients.

\begin{table}[h]
    \caption{Correlations between effectiveness of the apology representations and post-trust-violation emotional response}
    \label{tab:emoCorr}
    \begin{tabular}{ld{-1}d{-1}d{-1}}
        \toprule
        & \multicolumn{1}{c}{Low-criticality} 
        & \multicolumn{1}{c}{Mid-criticality} 
        & \multicolumn{1}{c}{High-criticality}\\
        \midrule
        Post-trust-violation emotional response
        & 0.248^*      &0.260^{***}  &0.332^{***} \\
        \bottomrule
        \multicolumn{4}{l}{Significance codes: $ ^* p < 0.05, ^{**} p < 0.01, ^{***} p < 0.001$}
    \end{tabular}
\end{table}

\subsection{Types of Apology Representations and Emotional Response}
Although not a part of the prior hypotheses, we are also interested in whether the types of apology representations make a difference in participants’ emotional responses after they are presented with the apology message. To examine this impact, the post-repair emotional response scores for all three trust violation scenarios were subjected to a one-way ANOVA with factor different apology representations (\textit{text}, \textit{emoji}, and \textit{cartoon car}). The ANOVA results revealed that only in the high-criticality scenario, there exists a difference in participants’ emotional response among different apology representations $(F(2, 1071) = 3.262, p < 0.05, \eta^2 = 0.006)$. A post hoc Tukey HSD test showed that \textit{cartoon car} $(M = -1.5, SD = 2.71)$ and \textit{emoji} $(M = -2.01, SD = 2.7)$ differed significantly at $p < 0.05$; \textit{text} $(M = -1.71, SD = 2.7)$ was not significantly different from the other two permutations and fell somewhere in the middle.

\section{Discussion}
This study investigates the effectiveness of different anthropomorphic cues that express emotion in repairing trust and the extent to which individual differences and a person’s post-trust-violation emotional response may impact the effectiveness of a trust repair strategy. Overall, our results found the differences among anthropomorphic cues with emotional expression to be non-significant. Some personality traits (\textit{Openness} and \textit{Conscientiousness}) have a negative correlation with the effectiveness of the apologies. Age was also identified as a relevant factor in trust repair. Moreover, we found a positive correlation between a person’s post-trust-violation emotional response and the effectiveness of the apology messages.

We will discuss our findings concerning each of the hypotheses formulated for this study.

\subsection{Non-significance of effects from Anthropomorphic Cues}
For our H1 (“\textit{An apology message featuring anthropomorphic cues that are emotionally expressive will impact how effective a trust repair strategy is.}”), we found no significant differences among the apology representations (\textit{text}, \textit{emoji}, and \textit{cartoon car}). This result may be due to our experiment setup, in which we utilized repeated exposures and the ordinary day-to-day scenario, which may have reduced our manipulation power.

In evaluating the participants’ free responses, we observed a polarizing perception of the anthropomorphic cues presented in this study, similar to prior studies that incorporate a testing design featuring emojis \cite{Grover_2020}. Some participants had very positive responses to the emojis; for example, one participant (P601: \textit{low criticality} × \textit{emoji}) stated “\textit{apparently I feel better when the device is humanized with a face and personal pronouns!}”, and another participant (P155: \textit{mid criticality} × \textit{emoji}) stated “\textit{it provided a `visual' sad face and gave me a proactive estimation without my having to push more buttons.}” Some other participants expressed different opinions; for example, one participant (P1255: \textit{high criticality} × \textit{cartoon car}) expressed the apology message as “\textit{...super annoying, especially if it can't help me figure out where to go. Seeing a sad car is even more annoying.}” Another participant (P511: \textit{high criticality} - \textit{emoji}) stated that “\textit{it should check or show you pictures of your destination. The emoji doesn't help.}”

Notably, some participants also expressed a rather negative attitude toward a GPS navigation system apologizing to them. For example, one participant (P349) stated “\textit{If I'm paying for a device to be convenient and fast and knowledgeable, why would I want to have to keep telling it its doing a poor job, then have it apologize to me like a sad robot? I just don't like the system's responses. It makes me feel like I hurt its feelings.}” Another participant (P957: \textit{mid criticality} × \textit{emoji}) stated that “\textit{I think the apologetic updated traffic time message would make me more upset than no message at all… rather than apologize for it I'd prefer to have the GPS just change the time to destination as it gets the correct information.}” Another participant (P323: \textit{high criticality} × \textit{emoji}) stated that “\textit{... The `apology' only makes it worse - it's the equivalent of `My Bad' - doubly offensive.}”

\subsection{Individual Differences}
For our H2 (“\textit{A participant’s personality impacts the effectiveness of a trust repair strategy.}”), we have found a negative correlation between the effectiveness of the apology message and a person’s \textit{openness} and \textit{conscientiousness}. We compare our findings with those of Chien et al.\cite{Chien_2016}, who measured a dispositional trust toward automation in general, and those of Rossi et al. \cite{Rossi_2018}, who measured an initial trust level toward a robot. It appears that \textit{conscientiousness} is the one dimension that is relevant across the board. \textit{Openness} is, however, a factor that did not present significant differences in prior studies.

The negative correlation may seem unintuitive and inconsistent with previous findings at first. However, the trust level measured in this study is a learned trust level that would fluctuate during interaction \cite{Hoff_2015}. Our data have also shown a positive correlation between openness and conscientiousness with the participants’ initial trust level. The result suggests that while openness and conscientiousness may cause a person to trust the automation more generally, this effect may not apply to a specific trust violation and trust repair attempt. This difference has also provided evidence that it is critical to distinguish the type of trust that is measured, and that trust repair is a unique context that requires special attention.

In addition to personality, we also found that age is a factor influencing the effectiveness of the apology messages. In particular, the apology messages seem to be less effective in repairing trust for older adults. Our finding is similar to Sanchez et al. \cite{Sanchez_2004}, who found that older adults were more sensitive in calibrating their trust level when perceiving a difference in the automation’s performance. Further studies are needed to understand whether some factors can be a more direct predictor. For instance, prior research has found a negative correlation between age and affinity to technology \cite{Franke_2019}.

\subsection{Emotional Response in the Trust Repair Process}
For our H3 (“\textit{The participant’s post-trust-violation emotional response will impact the effectiveness of a trust repair strategy.}”), our result shows a similarity to studies conducted in the organizational trust repair context \cite{Lockey_2017}, that there is a positive correlation between a person’s emotional response toward a trust violation and the effectiveness of the trust repair strategy. 

Our study setup did not allow us to draw a direct, consequential relationship between the emotional response and the effectiveness of the trust repair strategy. However, our findings have suggested that there are opportunities for an automation system to try mitigating the negative emotion prior or concurrent to the happening of trust repair to increase its effectiveness, as Tomlinson and Mayer’s trust model has conceptualized \cite{Tomlinson_2009}, especially with developments of emotionally intelligent agents in the future. 

\subsection{Limitations and Future Studies}
There are several limitations to this study that must be noted. 

First, this experiment was conducted in a scenario-based fashion, where the participants were prompted to read about and imagine themselves in certain scenarios instead of experiencing them. This setup has also limited us in measuring only the self-reported trust level, in contrast to behavior trust, for which other research has found incongruence between the two types of trust \cite{Kulms_2019}. We, therefore, encourage future studies to replicate the conditions in a Wizard-of-Oz or real-world setting.

Second, due to the exploratory nature of this study, we included a few different dimensions and employed a repeated design in the experiment. We favored the use of briefer forms of measurement, such as trust level, personality, and emotional response, to avoid survey fatigue, especially considering the unmoderated setup. Future studies should consider a narrower focus, more controlled environment, and more thorough measurement to increase the finding’s validity.

In conclusion, this study serves as an initial step in a series of explorations into individual differences and the role of emotion in human-automation interaction, particularly in the trust repair context. Despite the limitations, the data has shown differences in the effectiveness of trust repair strategies in regards to different personality traits and post-trust-violation emotional responses. 

\section*{Acknowledgments}
This work was funded by Toyota Motor Corporation as part of a long-term investigation of emotion in regards to human-machine interaction. We also thank their UX Design group for their valuable suggestions and discussions during this research project.

\bibliographystyle{unsrt} 
\bibliography{references}

\begin{thebibliography}{10}

\bibitem{Lee_See_2004}
John~D. Lee and Katrina~A. See.
\newblock Trust in automation: Designing for appropriate reliance.
\newblock {\em Human Factors}, 46(1):50--80, 2004.

\bibitem{Merritt_2008}
Stephanie~M. Merritt and Daniel~R. Ilgen.
\newblock Not all trust is created equal: Dispositional and history-based trust
  in human-automation interactions.
\newblock {\em Human Factors}, 50(2):194--210, 2008.

\bibitem{deVisser_2018}
Ewart~J. de~Visser, Richard Pak, and Tyler~H. Shaw.
\newblock From ‘automation’ to ‘autonomy’: the importance of trust
  repair in human–machine interaction.
\newblock {\em Ergonomics}, 61(10):1409--1427, 2018.

\bibitem{Hoff_2015}
Kevin~Anthony Hoff and Masooda Bashir.
\newblock Trust in automation: Integrating empirical evidence on factors that
  influence trust.
\newblock {\em Human Factors}, 57(3):407--434, 2015.

\bibitem{Schaefer_2016}
Kristin~E. Schaefer, Jessie Y.~C. Chen, James~L. Szalma, and P.~A. Hancock.
\newblock A meta-analysis of factors influencing the development of trust in
  automation: Implications for understanding autonomy in future systems.
\newblock {\em Human Factors}, 58(3):377--400, 2016.

\bibitem{Kraus_2020_a}
Johannes Kraus, David Scholz, Dina Stiegemeier, and Martin Baumann.
\newblock The more you know: Trust dynamics and calibration in highly automated
  driving and the effects of take-overs, system malfunction, and system
  transparency.
\newblock {\em Human Factors}, 62(5):718--736, 2020.

\bibitem{Stokes_2010}
Charlene~K. Stokes, Joseph~B. Lyons, Kenneth Littlejohn, Joseph Natarian, Ellen
  Case, and Nicholas Speranza.
\newblock Accounting for the human in cyberspace: Effects of mood on trust in
  automation.
\newblock In {\em 2010 International Symposium on Collaborative Technologies
  and Systems}, pages 180--187, 2010.

\bibitem{Goya-Martinez_2016}
Mariana Goya-Martinez.
\newblock Chapter 8 - the emulation of emotions in artificial intelligence:
  Another step into anthropomorphism.
\newblock In Sharon~Y. Tettegah and Safiya~Umoja Noble, editors, {\em Emotions,
  Technology, and Design}, Emotions and Technology, pages 171--186. Academic
  Press, San Diego, 2016.

\bibitem{Baker_2018}
Anthony~L. Baker, Elizabeth~K. Phillips, Daniel Ullman, and Joseph~R. Keebler.
\newblock Toward an understanding of trust repair in human-robot interaction:
  Current research and future directions.
\newblock {\em ACM Transactions on Interactive Intelligent Systems}, 8(4),
  November 2018.

\bibitem{Kim_2004}
Peter~H. Kim, Donald~L. Ferrin, Cecily~D. Cooper, and Kurt~T. Dirks.
\newblock Removing the shadow of suspicion: The effects of apology versus
  denial for repairing competence- versus integrity-based trust violations.
\newblock {\em Journal of Applied Psychology}, 89(1):104–118, 2004.

\bibitem{Kohn_2018}
Spencer~C. Kohn, Daniel Quinn, Richard Pak, Ewart~J. de~Visser, and Tyler~H.
  Shaw.
\newblock Trust repair strategies with self-driving vehicles: An exploratory
  study.
\newblock {\em Proceedings of the Human Factors and Ergonomics Society Annual
  Meeting}, 62(1):1108--1112, 2018.

\bibitem{Quinn_2018}
Daniel~B. Quinn.
\newblock Exploring the efficacy of social trust repair in human-automation
  interactions.
\newblock Master's thesis, Clemson University, Clemson, SC, May 2018.

\bibitem{Kim_2006}
Peter~H. Kim, Kurt~T. Dirks, Cecily~D. Cooper, and Donald~L. Ferrin.
\newblock When more blame is better than less: The implications of internal vs.
  external attributions for the repair of trust after a competence- vs.
  integrity-based trust violation.
\newblock {\em Organizational Behavior and Human Decision Processes},
  99(1):49--65, 2006.

\bibitem{Buchholz_2017}
Victoria Buchholz, Philipp Kulms, and Stefan Kopp Prof.~Dr.
\newblock It's (not) your fault! blame and trust repair in human-agent
  cooperation.
\newblock {\em Kognitive Systeme}, 2017(1), September 2017.

\bibitem{Jensen_2019}
Theodore Jensen, Yusuf Albayram, Mohammad Maifi~Hasan Khan, Md~Abdullah~Al
  Fahim, Ross Buck, and Emil Coman.
\newblock The apple does fall far from the tree: User separation of a system
  from its developers in human-automation trust repair.
\newblock In {\em Proceedings of the 2019 on Designing Interactive Systems
  Conference}, Dis '19, page 1071–1082, New York, NY, USA, 2019. Association
  for Computing Machinery.

\bibitem{Miramontes_2015}
Adriana Miramontes, Andriana Tesoro, Yuri Trujillo, Edward Barraza, Jillian
  Keeler, Alexander Boudreau, Thomas~Z. Strybel, and Kim-Phuong~L. Vu.
\newblock Training student air traffic controllers to trust automation.
\newblock {\em Procedia Manufacturing}, 3:3005--3010, 2015.

\bibitem{Chien_2016}
Shih-Yi Chien, Katia Sycara, Jyi-Shane Liu, and Asiye Kumru.
\newblock Relation between trust attitudes toward automation, hofstede’s
  cultural dimensions, and big five personality traits.
\newblock {\em Proceedings of the Human Factors and Ergonomics Society Annual
  Meeting}, 60(1):841--845, 2016.

\bibitem{Rossi_2018}
Alessandra Rossi, Kerstin Dautenhahn, Kheng~Lee Koay, and Michael~L. Walters.
\newblock The impact of peoples’ personal dispositions and personalities on
  their trust of robots in an emergency scenario.
\newblock {\em Paladyn, Journal of Behavioral Robotics}, 9(1):137--154, 2018.

\bibitem{Lockey_2017}
Steven~J. Lockey.
\newblock {\em The Role of Emotions and Individual Differences in the Trust
  Repair Process}.
\newblock Ph.d. dissertation., Durham University, Durham, UK, 2017.

\bibitem{Bagdasarov_2019}
Zhanna Bagdasarov, Shane Connelly, and James~F. Johnson.
\newblock Denial and empathy: Partners in employee trust repair?
\newblock {\em Frontiers in Psychology}, 10:19, 2019.

\bibitem{Tomlinson_2009}
Edward~C. Tomlinson and Roger~C. Mayer.
\newblock The role of causal attribution dimensions in trust repair.
\newblock {\em Academy of Management Review}, 34(1):85--104, 2009.

\bibitem{Merritt_2011}
Stephanie~M. Merritt.
\newblock Affective processes in human–automation interactions.
\newblock {\em Human Factors}, 53(4):356--370, 2011.

\bibitem{Kraus_2020_b}
Johannes Kraus, David Scholz, and Martin Baumann.
\newblock What’s driving me? exploration and validation of a hierarchical
  personality model for trust in automated driving.
\newblock {\em Human Factors}, page 0018720820922653, July 2020.

\bibitem{Waytz_2010}
Adam Waytz, John Cacioppo, and Nicholas Epley.
\newblock Who sees human?: The stability and importance of individual
  differences in anthropomorphism.
\newblock {\em Perspectives on Psychological Science}, 5(3):219--232, 2010.

\bibitem{Gambino_2020}
Andrew Gambino, Jesse Fox, and Rabindra~A Ratan.
\newblock Building a stronger casa: extending the computers are social actors
  paradigm.
\newblock {\em Human-Machine Communication}, 1:71--86, 2020.

\bibitem{Waytz_2014}
Adam Waytz, Joy Heafner, and Nicholas Epley.
\newblock The mind in the machine: Anthropomorphism increases trust in an
  autonomous vehicle.
\newblock {\em Journal of Experimental Social Psychology}, 52:113--117, 2014.

\bibitem{Kulms_2019}
Philipp Kulms and Stefan Kopp.
\newblock More human-likeness, more trust? the effect of anthropomorphism on
  self-reported and behavioral trust in continued and interdependent
  human-agent cooperation.
\newblock In {\em Proceedings of Mensch Und Computer 2019}, MuC'19, page
  31–42, New York, NY, USA, 2019. Association for Computing Machinery.

\bibitem{Song_2020}
Yao Song and Yan Luximon.
\newblock Trust in ai agent: A systematic review of facial anthropomorphic
  trustworthiness for social robot design.
\newblock {\em Sensors}, 20(18), 2020.

\bibitem{deVisser_2016}
Ewart~J. de~Visser, Samuel~S. Monfort, Ryan McKendrick, Melissa A.~B. Smith,
  Patrick~E. McKnight, Frank Krueger, and Raja Parasuraman.
\newblock Almost human: Anthropomorphism increases trust resilience in
  cognitive agents.
\newblock {\em Journal of Experimental Psychology: Applied}, 22(3):331–349,
  September 2016.

\bibitem{Glikson_2020}
Ella Glikson and Anita~Williams Woolley.
\newblock Human trust in artificial intelligence: Review of empirical research.
\newblock {\em Academy of Management Annals}, 14(2):627--660, 2020.

\bibitem{Verberne_2015}
Frank M.~F. Verberne, Jaap Ham, and Cees J.~H. Midden.
\newblock Trusting a virtual driver that looks, acts, and thinks like you.
\newblock {\em Human Factors}, 57(5):895--909, 2015.

\bibitem{Matsui_2019}
Tetsuya Matsui and Seiji Yamada.
\newblock Designing trustworthy product recommendation virtual agents operating
  positive emotion and having copious amount of knowledge.
\newblock {\em Frontiers in Psychology}, 10:675, 2019.

\bibitem{Mishler_2019}
Scott Mishler.
\newblock Whose drive is it anyway? using multiple sequential drives to
  establish patterns of learned trust, error cost, and non-active trust repair
  while considering daytime and nighttime differences as a proxy for
  difficulty.
\newblock Master's thesis, Old Dominion University, Norfolk, VA, 2019.

\bibitem{Lee_Moray_2004}
John~D. Lee and Neville Moray.
\newblock Trust, self-confidence, and operators' adaptation to automation.
\newblock {\em International Journal of Human-Computer Studies},
  40(1):153--184, 1994.

\bibitem{Lewicki_2017}
Roy~J. Lewicki and Chad Brinsfield.
\newblock Trust repair.
\newblock {\em Annual Review of Organizational Psychology and Organizational
  Behavior}, 4(1):287--313, 2017.

\bibitem{Gosling_2003}
Samuel~D Gosling, Peter~J Rentfrow, and William~B Swann.
\newblock A very brief measure of the big-five personality domains.
\newblock {\em Journal of Research in Personality}, 37(6):504--528, 2003.

\bibitem{McKnight_2002}
D.~Harrison McKnight, Vivek Choudhury, and Charles Kacmar.
\newblock Developing and validating trust measures for e-commerce: An
  integrative typology.
\newblock {\em Information Systems Research}, 13(3):334--359, 2002.

\bibitem{Watson_1998}
David Watson, Lee~Anna Clark, and Auke Tellegen.
\newblock Development and validation of brief measures of positive and negative
  affect: The panas scales.
\newblock {\em Journal of Personality and Social Psychology},
  54(6):1063–1070, 1988.

\bibitem{Garcia_2016}
Sarah~E. Garcia and Laura~M. Hammond.
\newblock Capturing \& measuring emotions in ux.
\newblock In {\em Proceedings of the 2016 CHI Conference Extended Abstracts on
  Human Factors in Computing Systems}, Chi Ea '16, page 777–785, New York,
  NY, USA, 2016. Association for Computing Machinery.

\bibitem{Grover_2020}
Ted Grover, Kael Rowan, Jina Suh, Daniel McDuff, and Mary Czerwinski.
\newblock Design and evaluation of intelligent agent prototypes for assistance
  with focus and productivity at work.
\newblock In {\em Proceedings of the 25th International Conference on
  Intelligent User Interfaces}, Iui '20, page 390–400, New York, NY, USA,
  2020. Association for Computing Machinery.

\bibitem{Sanchez_2004}
Julian Sanchez, Arthur~D. Fisk, and Wendy~A. Rogers.
\newblock Reliability and age-related effects on trust and reliance of a
  decision support aid.
\newblock {\em Proceedings of the Human Factors and Ergonomics Society Annual
  Meeting}, 48(3):586--589, 2004.

\bibitem{Franke_2019}
Thomas Franke, Christiane Attig, and Daniel Wessel.
\newblock A personal resource for technology interaction: Development and
  validation of the affinity for technology interaction (ati) scale.
\newblock {\em International Journal of Human–Computer Interaction},
  35(6):456--467, 2019.

\end{thebibliography}

\end{document}